\documentstyle[12pt]{article}
\newcommand{\beq}{\begin{equation}}
\newcommand{\eeq}{\end{equation}}
\newcommand{\eq}[1]{(\ref{#1})}
\newcommand{\beqn}{\begin{eqnarray}}
\newcommand{\eeqn}{\end{eqnarray}}
\newcommand{\dst}{&\displaystyle}
\newcommand{\al}{\mbox{$Z\alpha$}}
\newcommand{\alp}{\mbox{$\alpha$}}
\newcommand{\eps}{\mbox{$\varepsilon$}}
\newcommand{\Q}{\mbox{$\kappa$}}
\newcommand{\r}{\mbox{$\vec{r}$}}
\newcommand{\ri}{\mbox{$\vec{r}_1$}}
\newcommand{\rii}{\mbox{$\vec{r}_2$}}
\newcommand{\g}{\mbox{$\vec{\gamma}$}}
\newcommand{\rp}{\mbox{$\vec{r}\,'$}}
\newcommand{\val}{\mbox{$\vec{\alpha}$}}
\newcommand{\n}{\mbox{$\vec{n}$}}
\newcommand{\ei}{\mbox{$\vec{e}_1$}}
\newcommand{\eii}{\mbox{$\vec{e}_2$}}
\newcommand{\bi}[1]{\bibitem{#1}}
\newcommand{\fr}[2]{\frac{#1}{#2}}
\newcommand{\np}{\mbox{$\vec{n}\,'$}}
\newcommand{\pv}{\mbox{$\vec{p}$}}
\newcommand{\ki}{\mbox{$\vec{k}_1$}}
\newcommand{\kii}{\mbox{$\vec{k}_2$}}
\newcommand{\k}{\mbox{$\vec{k}$}}
\begin{document}

\begin{flushright}BUDKERINP 94-97  \\ December 1994
\end{flushright}

\vspace{1.0cm}
\begin{center}{\Large \bf Quasiclassical Green function
and Delbr\"uck scattering in a screened Coulomb field}\\
\vspace{1.0cm}

{\bf  R.N.~Lee and A.I.~Milstein} \\
G.I.~Budker Institute of Nuclear Physics,\\
630090 Novosibirsk, Russia

\vspace{4.0cm}
\end{center}

\begin{abstract}

A simple integral representation is derived for the quasiclassical Green
function of the Dirac equation in an arbitrary spherically-symmetric
decreasing external field. The consideration is based on
the use of the  quasiclassical radial wave functions
with the main contribution of large angular momenta taken into account.
The Green function obtained is applied to the calculation of the
Delbr\"uck scattering amplitudes in a screened Coulomb field.
\end{abstract}

\newpage

\section{\bf Introduction}

The most convenient way to take into account the external electromagnetic
field in quantum electrodynamic processes is the use of the Furry
representation. So, it is necessary to know the Green function
$G(\r,\rp |\eps)$  of the  Dirac equation in this field. Unfortunately,
the explicit forms of the Green functions are known only for the few
potentials and numerical calculations should be exploited.
For many  high-energy QED processes the main contribution to the
amplitudes is provided by large angular momenta.
Therefore, one can use the quasiclassical approximation.
In the present paper, the explicit expression of the  quasiclassical Green
function of the Dirac equation in an arbitrary spherically-symmetric
decreasing external field is found. Previously the quasiclassical Green
function of the Dirac equation has been obtained in \cite{MS1,MS2} for the
case of the Coulomb field by summing the integral representation of the
exact Green function \cite{MS3} over large angular momenta.
As it will be shown, to obtain the quasiclassical Green function,
it is not necessary to know  the exact one.
It is sufficient to use  the  quasiclassical radial wave functions
at large angular momenta. This method has been applied earlier
in \cite{OMW} to derive Sommerfeld-Maue type wave functions \cite{ZM}
used at the consideration of high-energy bremsstrahlung and pair production
in a screened Coulomb field. The integral representation of the Green
function obtained in our paper is convenient in analytic calculations
of the amplitudes of different high-energy QED processes in the
external field. To confirm this statement we calculate the Delbr\"uck
scattering amplitude  \cite{D} (the elastic scattering of a photon in the
external field via virtual electron-positron pairs) in a screened Coulomb
field.

Delbr\"uck scattering is one of the few nonlinear QED processes which can
be tested by experiment with high accuracy (see recent review \cite{MShu}).
At the present time Delbr\"uck amplitudes are studied in detail in the
Coulomb field exactly in the parameter  $\al$ at high photon energy
$\omega \gg m$  only;  $m$ is the electron mass,
 $Z|e|$ is the charge of the nucleus, $\al= e^2 = 1/137 $ is the
fine-structure constant, $e$ is the electron charge, $\hbar =c=1$.
The approaches used essentially depended on  the momentum transfer
$\Delta= |\kii -\ki |$ ( $ \ki$ and $ \kii$ being the
momenta of the incoming and outgoing photons, respectively).
At $\Delta\ll \omega $ the amplitudes have been found in \cite{CW1,CW2,CW3}
by summing in a definite approximation the Feynman diagrams with an
arbitrary number of photons exchanged with a Coulomb centre, and also in
\cite{MS1,MS2} with the help of the  quasiclassical Green function in a
Coulomb field.
At $m\ll\Delta\sim\omega$ the amplitudes of the process have been obtained
in \cite{MSha,MRS1,MRS2} using the exact electron Green function
in a Coulomb field \cite{MS3} in the limit $m=0.$
Many authors  have performed the calculations for an arbitrary
photon energy  $\omega $ but only in the lowest-order Born approximation with
respect to the parameter $\al$  (the results obtained in this approximation
are surveyed in \cite{PM}).  It turned out that Coulomb corrections at
$\al \sim 1$ and  $\omega\gg m$ drastically change the result as compared
to the Born approximation.

The effect of screening is important only in the case of small
momentum transfer $\Delta \sim 1/r_{c}\ll m$, where $r_{c}$  is the
screening radius of the atom.  It is this range of momentum transfer that
we consider in our paper.

\section{\bf Green function}

Let us consider the Green function of the Dirac equation in the external
spherically-symmetric field $V(r)$ :
\beq\label{g1}
G(\r,\rp |\,\eps )= \,\fr{1}{\gamma^{0}(\eps\, -\, V(r))\,
-\,\g\pv\, -\, m +i0}\,\delta (\r -\rp )\, ,
\eeq
where $\gamma^{\mu}$ are Dirac matrices, $\pv =-i\vec{\nabla}$ .  We are
interested in the calculation of the Green function at $|\eps |\gg m.$
Let us represent the function $G$ in the form
\beq\label{g2}
G(\r,\rp |\,\eps )= \left[ \gamma^{0}(\eps\, -\, V(r))\,
-\,\g\pv\, +\, m \right] D(\r,\rp |\,\eps )\, ,
\eeq
where the function $D(\r,\rp |\,\eps )$ is
\beq\label{g3}
D(\r,\rp |\,\eps )= \,\fr{1}{(\eps\, -\, V(r))^2\,
-\,\pv\,^2\, -[\val\pv,V(r)] \, -\, m^2+i0}\,\delta (\r -\rp )\, .
\eeq
Here $\val = \gamma^0\g$ .
As it is known (see \cite{ZM}), at high energies $\eps\gg m$ one can
neglect $V^2(r)$ in \eq{g3} and take into account  only the first term
of the expansion with respect to the commutator $[\val\pv,V(r)]$.
Making the cited expansion and using the representation
\beq\label{H}
[\val\pv,V(r)]\, =\,\fr{1}{2\eps}[\val\pv,H]\quad ,\quad
H=\pv\,^2 + 2\eps V(r)  \, ,
\eeq
we get the following representation for the function $D$ :
\beq\label{D}
D(\r,\rp |\,\eps )= \,\left[1\, -\,\fr{i}{2\eps}(\val , \vec{\nabla} +
\vec{\nabla}\,'\,)\right]D^{(0)}(\r,\rp |\,\eps )\, ,
\eeq
where
\beq\label{D0}
D^{(0)}(\r,\rp |\,\eps )= \,\fr{1}{\Q^2\, -\,  H\, +i0}\,\delta (\r -\rp )\, ,
\eeq
$\Q^2=\eps^2-m^2$ . Thus, the problem reduces to the calculation of the
quasiclassical Green function $D^{(0)}$ of the Schr\"odinger equation
with the hamiltonian $H.$

Let us introduce the impact parameter $\rho = |\r\times\rp|/|\r - \rp |$.
In high-energy processes the characteristic distances are
$|\r - \rp |\sim \Q /m^2 \gg 1/m $  and $\rho \geq 1/m$ .
So, the corresponding angular momentum $l\sim \Q\rho \gg 1$, and one can
use the quasiclassical approximation. Besides, we consider the case
$\rho \ll |\r - \rp|$ . Hence,  the angle either between
$\r$ and $-\rp$ or between $\r$ and $\rp$ is small.

Consider the set of eigenfunctions of the hamiltonian $H$ and use in
\eq{D0} its completeness, replacing $\delta$-function by the sum of
products of eigenfunctions. Obviously, the main contribution to
$D^{(0)}$ is provided by the functions of the continuous spectrum
with large angular momentum values. The eigenfunction
$\psi_{\vec{q}}(\r)$, containing  the plane wave with the
momentum $\vec{q}$ and the outgoing  spherical wave in its asymptotic form,
can be represented as:
\beq\label{psi}
\psi_{\vec{q}}(\r)=\fr{1}{qr}\sum_{l=0}^{\infty}\,
i^{l}\mbox{e}^{i\delta_{l}}(2l+1)u_{l}(r)P_{l}(\cos \vartheta)\, .
\eeq
Here $P_{l}(x)$ are the Legendre polynomials, $\vartheta$ is the angle
between vectors $\vec{q}$ and $\r$. The set of eigenfunctions with the
ingoing spherical waves in asymptotics leads to the same result for the
Green function. In the quasiclassical approximation the functions $u_{l}(r)$
and $\delta_{l}=\delta(l/q)$  are equal to (see \cite{OMW}) :
\beqn\label{u}
u_{l}(r)=\sin \left(qr-l\pi /2+l^2/2qr+\lambda\delta(l/q)+\lambda\Phi(r)
\right) \, , \\
\Phi(r)=\int\limits_{r}^{\infty}\, V(\zeta)d\zeta\quad ,\quad
\delta(\rho)= -\int\limits_{0}^{\infty}\,
V\left(\sqrt{\zeta^2+\rho^2}\right)d\zeta
\quad , \quad \lambda=\eps /q \, . \nonumber
\eeqn
Taking into account the  completeness relation, we get:
\beq\label{D1}
D^{(0)}(\r,\rp |\,\eps )=
\int\limits \,\fr{\psi_{\vec{q}}(\r)\psi^{*}_{\vec{q}}(\rp)}
{\Q^2\, -\,  q^2\, +i0}\,\fr{d\vec{q}}{(2\pi)^3}\, .
\eeq
Substituting \eq{psi} into \eq{D1}  and taking the integral over the
angles of vector  $\vec{q}$ with the help of well-known relation for the
Legendre polynomials
$$
\int\limits P_{l}(\vec{n}_1\vec{n}_2)P_{l'}(\vec{n}_1\vec{n}_3)\, d\Omega_1\,=
\fr{4\pi}{2l+1}P_{l}(\vec{n}_2\vec{n}_3)\delta_{ll'} \, ,
$$
where $\vec{n}_{i}$ are unit vectors, we obtain:
\beq\label{D2}
D^{(0)}(\r,\rp |\,\eps )= \fr{1}{2\pi^2\, rr'}\int\limits_{0}^{\infty}\,\fr{dq}
{\Q^2\, -\,  q^2\, +i0}\sum_{l=0}^{\infty}\,
(2l+1)u_{l}(r)u_{l}(r')P_{l}(\n\np)\, ,
\eeq
$\n=\r/r$, $\np =\rp /r'$ . Using \eq{u}, we represent the product
$u_{l}(r)u_{l}(r')$ as follows :
\beqn\label{uu}
\dst
u_{l}(r)u_{l}(r')=\fr{1}{2}\cos\left[q(r-r')+l^2(r'-r)/2qrr'+\lambda
(\Phi(r)-\Phi(r'))\right] \,  \\
\dst
-\,\fr{1}{2}(-1)^{l}\cos\left[q(r+r')+l^2(r+r')/2qrr'+2\lambda\delta(l/q)
+\lambda\left(\Phi(r)+\Phi(r')\right)\right] \, . \nonumber
\eeqn
If the angle $\theta$ between the vectors $\n$ and $-\np$ is small, then
one can replace $P_{l}(\n\np)$ by $(-1)^{l}J_0(l\theta),$
where $J_{\nu}(x)$ is the Bessel function. After this substitution it is
clear that  in sum over $l$ the main contribution comes from the second term in
\eq{uu}. For this term  the summation with respect to $l$ can be replaced
by an integration. Let us make in \eq{D2} the exponential parametrization
of the energy denominator :
$$
\fr{1}{\Q^2\, -\,
  q^2\, +i0}=-i\int\limits_{0}^{\infty}\exp[is(\Q^2\, -\, q^2\,)]ds
$$
Then the integrals over $q$ and $s$ can be taken by means of the stationary
phase method. After simple calculations for the case under consideration
$\theta\ll 1$, one obtains:
\beq\label{D0F}
D^{(0)}(\r,\rp |\,\eps )=\fr{i\mbox{e}^{i\Q(r+r')}}{4\pi\Q rr'}
\int\limits_{0}^{\infty}dl lJ_{0}(l\theta)\exp\biggl\{i\biggl[
\fr{l^2(r+r')}{2\Q rr'}+2\lambda\delta(l/\Q)+\lambda(\Phi(r)+\Phi(r'))
\biggr]\biggr\} \, .
\eeq
In this formula  $\lambda=\eps /\Q$. In the relativistic case
$\lambda=+1$ at $\eps >0$ and $\lambda=-1$ at $\eps <0$.

If the angle $\theta_1=\pi - \theta$ between vectors $\n$ and $\np$ is small
then one can replace $P_{l}(\n\np)$ by $J_0(l\theta_1).$  In this case
 the main contribution in the sum over $l$ comes from the first term in
\eq{uu}.
Since it doesn't contain $\delta(l),$ it is possible to take the integral over
$l$ after the transformations similar to those performed at the derivation of
\eq{D0F}. At $\pi -\theta\ll 1$ one has:
\beq\label{D0F1}
D^{(0)}(\r,\rp |\,\eps )= \,-\fr{1}{4\pi |\r-\rp|}\exp\left\{
i\Q|\r-\rp|\, +i\lambda\mbox{sign}(r-r')(\Phi(r)-\Phi(r'))\right\}
\quad .
\eeq
Substituting \eq{D0F} into \eq{D0F1} and \eq{D}, we find for the function $D$
at $ \theta\ll 1$ :
\beqn\label{DT}
\dst
D(\r,\rp |\,\eps )= \,\fr{i\mbox{e}^{i\Q(r+r')}}{4\pi\Q rr'}
\int\limits_{0}^{\infty}\,dl l\left[J_{0}(l\theta)-
i\fr{(\val ,\n+\np)}{\Q\theta}\delta^{\prime}(l/\Q)J_{1}(l\theta)
\right]  \nonumber \\
\dst
\times\exp\biggl\{i\biggl[l^2(r+r')/2\Q rr'+
2\lambda\delta(l/\Q)+\lambda(\Phi(r)+\Phi(r'))\biggr]\biggr\}\quad  .
\eeqn
Here $\delta^{\prime}(\rho)= \partial\delta(\rho) /\partial \rho.$

At $ \pi - \theta \ll 1$ the function $D$ is of the following form:
\beqn\label{DT1}
\dst
D(\r,\rp |\,\eps )=
-\left[1-\mbox{sign}(r-r')(V(r)-V(r'))(\val ,\n+\np)/4\Q\, \right]
\nonumber\\
\dst
\times\exp\left\{
i\Q|\r-\rp|\, +i\lambda\mbox{sign}(r-r')(\Phi(r)-\Phi(r'))\right\}/
{4\pi |\r-\rp|}\quad  .
\eeqn
Substituting \eq{DT} and \eq{DT1} into \eq{g2}, we get the final result
for the quasiclassical Green function of the Dirac equation in
spherically-symmetric external field. One has at $\theta\ll 1$
\beqn\label{GT}
\dst
G(\r,\rp |\,\eps )= \,\fr{i\mbox{e}^{i\Q(r+r')}}{4\pi\Q rr'}
\int\limits_{0}^{\infty}dl l\,
\exp\biggl\{i\biggl[l^2(r+r')/2\Q rr'+
2\lambda\delta(l/\Q)+\lambda(\Phi(r)+\Phi(r'))\biggr]\biggr\} \nonumber \\
\dst
\times\biggl\{\biggl[ \gamma^{0}\eps+m -\fr{1}{2} (\g ,\n-\np )
(\Q+l^2/2\Q rr')\biggr] J_{0}(l\theta)+i\biggl[ l^2(r-r')(\g,\n+\np)/2rr'  \\
\dst
+l\delta^{\prime}(l/\Q)\gamma^{0}\biggl(1-(\g,\n)(\g,\np)-
(\g,\n+\np)m/\Q\biggr)\biggr]J_{1}(l\theta)/(l\theta)\biggr\} \, ,
\nonumber
\eeqn
and at $ \pi - \theta \ll 1$
\beqn\label{GT1}
\dst
G(\r,\rp |\,\eps )=
-\fr{1}{4\pi R}[ \gamma^{0}\eps+m -(\Q+i/R)(\g,\vec{R})/R ] \nonumber \\
\dst
\times\exp\left\{i\Q R\, +i\lambda\mbox{sign}(r-r')(\Phi(r)-\Phi(r'))\right\}
 \quad , \quad  \vec{R}= \r -\rp \, .
\eeqn
In the Coulomb field $V(r)= -\al/r$, we have
\beq\label{del}
2\delta(\rho)\, +\, \Phi(r)+\Phi(r') = \al\ln (4rr'/\rho^2) \quad ,
 \quad \delta^{\prime}(\rho)=-\al/\rho \, .
\eeq

Using \eq{del} and \eq{GT}, we find that our result for the Green function
in the Coulomb field is in agreement with that obtained in \cite{MS1,MS2}.

\section{\bf Delbr\"uck scattering}

Let us apply the formulae obtained to the calculation of the
Delbr\"uck amplitudes in a screened Coulomb field.
In the Thomas-Fermi model the screening radius
$r_{c}\sim (m\alpha)^{-1}Z^{-1/3}$ .
The characteristic impact parameter $\rho\sim 1/\Delta.$  If
$R\ll 1/\Delta \ll r_{c}$ ($R$ is the radius of the nucleus), then
the  screening can be neglected and the amplitude under consideration
coincides with that in the Coulomb field. If $1/\Delta \sim r_{c}\gg 1/m,$
then it is necessary to take  screening into account.
For this momentum transfer  the main contribution to the amplitude is
provided by impact parameters $\rho$ from $1/m$ to $r_{c}.$
The corresponding angular momenta $l\sim \omega\rho \gg 1$ and
the quasiclassical approximation is valid.

Let an initial photon with momentum $\ki$ produce at the point
$\ri$ a pair of virtual particles which is transformed at the point $\rii$
into a photon with momentum $\kii$. Then the uncertainty relation
gives $\tau\sim |\rii -\ri | \sim \omega/(m^2+\Delta^2)$  for the lifetime
of the virtual pair. Therefore, at $\omega/m^2 \gg r_{c} $ the angles
between $\ki\, , \kii \, , \rii $ and $-\ri $ are small.
It is this energy range we consider further.
According to the Feynman rules, in the Furry representation  the
Delbr\"uck scattering amplitude reads
\beq\label{a1}
M=\,2i\alp\,\int\limits d\ri\,d\rii \exp[ i(\ki\ri-\kii\rii )]\,\int\limits
d\eps\, Tr\hat e_2^{*} G(\rii ,\ri |\omega -\eps)\hat e_1 G(\ri ,\rii |-\eps)
\; ,
\eeq
where  $e_1^{\mu}$ and $e_2^{\mu}$ are the polarization vectors of
initial and final photons, respectively, $\hat e = e_{\mu}\gamma^{\mu}.$
It is necessary to subtract, from the integrand for $M$ in \eq{a1},
the value of this integrand at zero potential. In the following such a
subtraction is assumed to be made and we perform it explicitly
in the final result. The main contribution to the amplitude $M$ arises
at the integration over $\eps$ from $m$ to $\omega -m.$ Thus,
 $\lambda= +1$ in the first Green function in \eq{a1} and  $\lambda= -1$
in the second one.  Using the representation \eq{g2} ,
it is convenient  to rewrite eq. \eq{a1} in the form
\beqn\label{a2}
\dst
M=i\alp\int\limits d\ri d\rii\exp [i(\ki\ri -\kii\rii)]\int\limits
d\eps\, Tr\biggl[(2\eii^{*}\pv_2-\hat e_2^{*}\hat k_2)
D(\rii ,\ri |\omega -\eps)\biggr]\\
\dst
\times\biggl[ (2\ei\pv_1+\hat e_1\hat k_1) D(\ri ,\rii |-\eps)\biggr]\, +
\,2i\alp\eii^{*}\ei\,\int\limits d\r\exp[ i(\ki -\kii)\r ]\,\int\limits
d\eps\, Tr D(\r ,\r |\eps) \; . \nonumber
\eeqn
Here $\pv_{1,2} =-i\vec{\nabla}_{1,2}$.
The last term in \eq{a2} doesn't contribute to the amplitude at high
energy because it is independent of $\omega$ and depends only on
momentum transfer $\Delta$. However, the amplitude at $\omega\gg \Delta$
is proportional to $\omega$ (see, i.g., \cite{MShu}).
Our further transformations are as follows. We substitute \eq{DT} into \eq{a2},
perform the differentiation and take the trace over $\gamma$-matrices.
It is convenient to direct the axis of the spherical coordinate system
along $\ki +\kii$ . In the small-angle approximation one has:
$d\Omega_{1,2}\approx\theta_{1,2}d\theta_{1,2}d\phi_{1,2}=
d\vec{\theta}_{1,2}$ with $(\vec{\theta}_{1,2},\ki +\kii )=0$ .
The Bessel functions in the Green function depend on the
 vectors $\vec{\theta}_{1,2}$ via the combination
$\theta = |\vec{\theta}_{1}+\vec{\theta}_{2}|$ only.
Let us change over to the variables
$\vec{\theta}=\vec{\theta}_{1}+\vec{\theta}_{2}$ and $\vec{\xi}=
r_1\vec{\theta}_{1}-r_2\vec{\theta}_{2}$. After that it is easy to take
the integral over $d\vec{\xi}$. Further, to demonstrate the method of
calculations, we consider the case of zero momentum transfer
($\kii =\ki =\k $), and present then the results of
similar calculations for $\Delta\sim 1/r_{c}$.

\subsection{\bf Zero momentum transfer}

We set $\ki =\kii ,\, \vec{e}_1 =\vec{e}_2$  and take the integral
with respect to $d\vec{\theta}$ with the help of the relation
(\cite{GR}, p. 732)
$$
\int\limits_{0}^{\infty}\, dxx\, e^{icx^2}J_{\nu}(ax)J_{\nu}(bx)=
\fr{ie^{i\pi\nu/2}}{2c}J_{\nu}(ab/2c)\exp\left[\fr{-i(a^2+b^2)}{4c}\right]\,
,
$$
and also those obtained by differentiating this expression with respect to
the parameters. Let us make the substitution of the variables
in the integral representation of the Green function:
$l_1=\Q_1\rho_1\, ,\, l_2=\Q_2\rho_2$ where $\Q_1=(\eps^2-m^2)^{1/2}$
,  $\Q_2=((\omega-\eps)^2-m^2)^{1/2}$ . After that it is convenient to pass
from the variables $r_1$ and $r_2$ to $s$ and $x$ :
$r_1=\Q_1\Q_2/[m^2\omega sx]$,
 $r_2=\Q_1\Q_2/[m^2\omega s(1-x)].$ As the result, the integral over $\eps$
becomes trivial and we get the following expression:
\beqn\label{a3}
\dst
M=\fr{2i\alp\omega m^2}{3}\,\int\limits_{0}^{1}\fr{dx}{x(1-x)}
\left[1+\fr{1}{x(1-x)}\right]
\int\limits_{0}^{\infty}\!\int\limits_{0}^{\infty}\,\rho_1\rho_2 d\rho_1
d\rho_2
\int\limits_{0}^{\infty}\fr{ds}{s} \\
\dst
\times\,\exp\biggl\{\fr{i}{2}\biggl[m^2(\rho_1^2+\rho_2^2)s -1/[sx(1-x)]\,
\biggr]\biggr\}\,\sin^2(\delta (\rho_2)-\delta (\rho_1))\,
 J_0(m^2s\rho_1\rho_2)\, , \nonumber
\eeqn
Here we have subtracted from the integrand its value at the field equal to
zero. Let us make the substitution of the variables
$\rho_1=\rho \mbox{e}^{-\tau/2}\, , \,
\rho_2=\rho \mbox{e}^{\tau/2}$ and deform the contour of the
integration with respect to $s$ so that the integral is extended
from zero to $i\infty$. Then the integral over $s$ can be taken using
the relation (\cite{GR}, p. 739 )
$$
\int\limits_{0}^{\infty}\exp[-x(a^2+b^2)/2\, -\, 1/2x\, ]I_{\nu}(abx)\fr{dx}{x}
=2I_{\nu}(a)K_{\nu}(b) \quad , \quad a<b\, ,
$$
where $I_{\nu}(x)$ and $K_{\nu}(x)$  are the modified Bessel functions of
the first and third kind respectively. As the result, we obtain
\beqn\label{a4}
\dst
M=\fr{8i\alp\omega m^2}{3}\,\int\limits_{0}^{1}\fr{dx}{x(1-x)}
\left[1+\fr{1}{x(1-x)}
\right]\int\limits_{0}^{\infty}\,\rho^3d\rho\,   \nonumber \\
\dst
\times\int\limits_{0}^{\infty}d\tau
\,\sin^2\biggl(\delta (\rho \mbox{e}^{\tau/2})-
\delta(\rho \mbox{e}^{-\tau/2})\biggr)\,I_0(y_1)K_0(y_2) \, ,
\eeqn
where $y_{1,2}=m\rho \mbox{e}^{\mp\tau/2}\,[x(1-x)]^{-1/2}$ .

We divide the integral over $\tau$ into two parts: from 0 to $\tau_0$
and from $\tau_0$ to $\infty$ , where $1\gg \tau_0 \gg 1/(mr_{c})$ .
Let us begin our calculations from the second domain.
In this domain the main contribution is given by the impact
parameters $\rho < r_{c}$,
and the field can be considered as the Coulomb one.
Integrating over $x$ , $\rho$, and , finally, over $\tau$, we get
\beqn\label{M2}
M_2=i\fr{28\alp\omega}{9m^2}\int\limits_{\tau_0}^{\infty}\,
d\tau\fr{\mbox{ch} \tau
\sin^2(\al\tau)}{\mbox{sh}^3 \tau}= \nonumber \\
\,-i\fr{28\alp\omega(\al)^2}{9m^2}
[\mbox{Re}\psi(1-i\al)+C+\ln 2\tau_0\, -\, 3/2] \, .
\eeqn
Here $\psi(x)=d\ln \Gamma(x)/dx$ , $C= 0.577...$ is the Euler constant.

In the first domain the difference
$\delta(\rho\mbox{e}^{\tau/2})-\delta(\rho\mbox{e}^{-\tau/2})$
is small, and it is possible to expand the integrand with
respect to this quantity. Therefore, this domain gives the contribution
to the amplitude in the Born approximation only.
It is convenient to divide the integral over $\rho$ into two parts:
from zero to $\rho_0 $ and from $\rho_0$ to $\infty$, where
$r_{c}\gg\rho_0\gg 1/(m\tau_0)$ .
In the integral from zero to $\rho_0$ the field can be considered again
as the Coulomb one,
and the integrals left  can be easily taken. The corresponding
contribution reads:
\beq\label{M11}
M_{11}=i\fr{28\alp\omega(\al)^2}{9m^2}
\left[\ln (m\tau_0\rho_0)\, +\, C\, -\, 11/21\right] \, .
\eeq
In the integral from $\rho_0$  to $\infty$ one can use the asymptotics
of the Bessel functions $I_0(x)$ and $K_0(x)$ at large $x$ and extend the
integration over $\tau$ up to infinity. The corresponding result is
\beq\label{M12}
M_{12}=\, i\fr{28\alp\omega}{9m^2}\,\int\limits_{\rho_0}^{\infty}
\, \rho \left(\fr{\partial\delta}{\partial\rho}\right)^2 \, d\rho\, .
\eeq
The sum of \eq{M2}, \eq{M11} and \eq{M12} is equal to
\beq\label{MT}
M=i\fr{28\alp\omega(\al)^2}{9m^2}\left[\ln(m\rho_0/2)+
(\al)^{-2}\int\limits_{\rho_0}^{\infty}\rho
\left(\fr{\partial\delta}{\partial\rho}\right)^2 d\rho
-\mbox{Re}\psi(1-i\al)+\fr{41}{42}\right]\, .
\eeq
At $\rho\ll r_{c}$ the integral in \eq{M12} is equal to $\ln(r_{c}/\rho_0)+A$,
where $A$ is some constant. Therefore, the amplitude $M$ in \eq{MT}
is independent of $\rho_0$ , and one can put, for instance,
$\rho_0=2/m.$ So, we have obtained the final result
for the forward  Delbr\"uck scattering amplitude for an arbitrary screened
potential. The explicit value of the constant depends on the form of the
potential.

Let us consider the case of the Moli\`ere potential \cite{M}, which
approximates the potential in the Thomas-Fermi model:
\beq\label{V}
V(r)= -\fr{\al}{r}\sum_{i=1}^{3}\,\alpha_{i}\mbox{e}^{-\beta_{i}r} \, ,
\eeq
where $\alpha_{1}=0.1$ ,  $\alpha_{2}=0.55$ , $\alpha_{3}=0.35$,
$\beta_{i}=\, mZ^{1/3}b_{i}/121$ , $b_{1}=6$ , $b_{2}=1.2$ , $b_{3}=0.3$ .
The corresponding scattering phase is equal to
\beq\label{drho}
\delta (\rho)=\, \al \sum_{i=1}^{3}\,\alpha_{i}K_0(\beta_{i}\rho)\, .
\eeq
Substituting this expression into \eq{MT}, we get the final result
for the forward  Delbr\"uck scattering amplitude in the Moli\`ere
potential:
\beq\label{MT1}
M=\, i\fr{28\alp\omega(\al)^2}{9m^2}\left[\ln(183Z^{-1/3})\, - \, C
-\,\mbox{Re}\psi(1-i\al)\, -\, \fr{1}{42}\,\right]\, .
\eeq
As known, the imaginary part of the forward scattering amplitude
of the photon is connected with the total cross section $\sigma$
of electron-positron pair production by the relation
$\sigma=\, \mbox{Im}\, M/\omega$.  Due to this relation, our formula
\eq{MT1} is in agreement with the result of \cite{DBM} for the total
cross section of pair production in a screened potential.
Note that the real part of the amplitude \eq{MT1} in the screened Coulomb
potential is equal to zero in contrast to the case of pure Coulomb
potential \cite{CW2,MS2}.

\subsection{\bf Non-zero momentum transfer}

At non-zero momentum transfer it is convenient to carry out the calculation
in terms of helicity amplitudes. One can choose the polarization vectors in
the form
\beq
\vec{e}_{1,2}^{\,\,\pm}\, =\, \bigl( [\vec{\lambda}\times\vec{\nu}_{1,2}]\, \pm
\,i\vec{\lambda} \bigr)/\sqrt{2}\; ,\quad
\vec{\lambda}=[\vec{\nu}_1\times\vec{\nu}_2 ]/
|[\vec{\nu}_1\times\vec{\nu}_2]| \, ,
\eeq
where $\vec{\nu}_{1,2}=\vec{k}_{1,2}/\omega $ .
There exist two independent amplitudes:
$M^{++}=M^{--}$ and $M^{+-}=M^{-+}$.
In terms of linear polarization, by virtue of parity conservation, the
amplitude differs from zero only when the polarization vectors of the
initial and final photon both lie in the scattering plane
 $(M^{\parallel})$ or are perpendicular to it $(M^{\perp})$.
These types of amplitudes are related via
$$
M^{\parallel}\,=\,M^{++}\,+\,M^{+-}\; ,\;
M^{\perp}\,=\,M^{++}\,-\,M^{+-}\, .
$$
At $\Delta =0$ the amplitude $M^{+-}$ vanishes by virtue of the
conservation of the angular momentum projection along the direction of
motion of the initial photon, and the amplitude  $M^{++}$
coincides with \eq{MT1} . Similar to the case of zero momentum transfer,
we divide the integral over $\tau$ into two parts: from 0 to $\tau_0$
and from $\tau_0$ to $\infty$ , where $1\gg \tau_0 \gg 1/(mr_{c}).$
The angle $\theta_0$ between $\ki$ and $\kii$ is
$\theta_0 = \Delta/\omega \ll m/\omega$ .
In the domain from $\tau_0$ to $\infty$ the field concides with the
Coulomb one and the angle $\theta_0$ can be neglected. The contribution of
this domain to the amplitude $M^{++}$ coincides with $M_2$ \eq{MT1} , and
the contribution to the amplitude $M^{+-}$ is equal to zero.
In the domain from zero to $\tau_0$ we split the integral over $\rho$
into two parts again: from zero to $\rho_0 $ and from $\rho_0$ to
$\infty$, where $r_{c}\gg\rho_0\gg 1/(m\tau_0)$. In the integral from
zero to $\rho_0 $ the field can be treated as a Coulomb one and the angle
$\theta_0$ can be neglected again. The corresponding contribution to
$M^{++}$ coincides with $M_{11}$ \eq{M11} , and the contribution to
$M^{+-}$ is equal to zero. The effect of screening is essential
in the last domain from $\rho_0 $ to $\infty$ only.
In this domain the main contribution to the integral over angles is given by
$\theta\sim \rho /r \sim \rho m^2/\omega \gg m/\omega \gg \theta_0$.
The argument of the Bessel functions in the expression for the Green
function is $l\theta\sim \omega\rho\theta\sim (m\rho)^2\gg 1,$ and
one can use the asymptotic expansion of the Bessel functions. It is
necessary to keep two terms of the expansion due to
the compensation. After that the integrals over $\theta$
and the other variables can be easily taken, and we
get the contribution of the domain under discussion to the amplitude
$M^{++}$ :
\beq\label{M12++}
M_{12}^{++}=\, i\fr{28\alp\omega}{9m^2}\,\int\limits_{\rho_0}^{\infty}
\, \rho \left(\fr{\partial\delta}{\partial\rho}\right)^2
\, J_0(\rho\Delta)\, d\rho\, .
\eeq
The sum of \eq{M12++}, \eq{M2} and \eq{M11} is:
\beq\label{MT++}
M^{++}=i\fr{28\alp\omega(\al)^2}{9m^2}\left[
\, (\al)^{-2}\int\limits_{2/m}^{\infty}\rho
\left(\fr{\partial\delta}{\partial\rho}\right)^2 J_0(\rho\Delta)d\rho\,
-\mbox{Re}\psi(1-i\al)+\fr{41}{42}\right]\, .
\eeq
Substitute \eq{drho} into \eq{MT++} and take the integral over $\rho$
by means of formula (6.578(10)) \cite{GR}. Then, the final result for the
amplitude $M^{++}$ in the case of Moli\`ere potential reads:
\beqn\label{MT1++}
\dst
M^{++}=\, i\fr{28\alp\omega(\al)^2}{9m^2}\biggl\{
-\,\mbox{Re}\psi(1-i\al)-C +\fr{41}{42}\, \\
\dst
-\fr{1}{2}\sum_{i,j}\alpha_{i}\alpha_{j}\left[\ln (\beta_{i}\beta_{j}/m^2)+
\fr{u}{\sqrt{u^2-1}}\ln(u+\sqrt{u^2-1})\right]\biggr\}\, ,\nonumber
\eeqn
where $u=(\Delta^2+\beta_{i}^2+\beta_{j}^2)/2\beta_{i}\beta_{j}$ .
At $\Delta \ll 1/r_{c}$ the formula \eq{MT1++} coincides with \eq{MT1}.
At $m\gg\Delta \gg 1/r_{c}$ the formula \eq{MT1++} turns into
\beq
M^{++}=\, i\fr{28\alp\omega(\al)^2}{9m^2}\left\{
\ln\fr{m}{\Delta}\, -\,\mbox{Re}\psi(1-i\al)-C +\fr{41}{42}\,\right\}\, ,
\eeq
in agreement with the result of \cite{CW1,CW2,MS2}.
Similarly, for the amplitude $M^{+-}$, we obtain:
\beq\label{MT+-}
M^{+-}=\, i\fr{4\alp\omega}{9m^2}\,\int\limits_{0}^{\infty}
\, \rho \left(\fr{\partial\delta}{\partial\rho}\right)^2
\, J_2(\rho\Delta)\, d\rho\, .
\eeq
Here the lower limit of the integral is replaced by zero
since the domain from zero to $\rho_0$ doesn't contribute.
For the Moli\`ere potential one has
\beqn\label{MT1+-}
\dst
M^{+-}=\, i\fr{2\alp\omega(\al)^2}{9m^2}\biggl\{
1\, +\fr{1}{\Delta^2}\sum_{i,j}\alpha_{i}\alpha_{j}\biggl[
(\beta_{i}^2-\beta_{j}^2)\ln\fr{\beta_{i}}{\beta_{j}}\, \\
\dst
-\,\fr{u(\beta_{i}^2+\beta_{j}^2)-2\beta_{i}\beta_{j}}
{\sqrt{u^2-1}}\ln(u+\sqrt{u^2-1})\biggr]\biggr\}\, . \nonumber
\eeqn
If $\Delta \rightarrow 0$ then $M^{+-}$ \eq{MT1+-} tends to zero. At
$m\gg\Delta \gg 1/r_{c}$ the formula \eq{MT1+-} turns into
\beq
M^{+-}=\, i\fr{2\alp\omega(\al)^2}{9m^2}\quad ,
\eeq
which is in accordance with the result of \cite{CW1,CW2,MS2}.

Thus, we have demonstrated on the example of  Delbr\"uck scattering
in a screened Coulomb potential that the quasiclassical Green function
obtained in the present paper can be used effectively at the consideration
of high-energy QED processes in an arbitrary  spherically-symmetric
decreasing external field.

We are grateful to V.M.Katkov and V.M.Strakhovenko for useful discussions.

\end{document}